\documentstyle[twoside,fleqn,espcrc2,psfig]{article}

\title{Hadron matrix elements for nucleon decay
       with the Wilson quark action
       \thanks{presented by N. Tsutsui}}

\author{JLQCD Collaboration:
        N. Tsutsui\address{Department of Physics, Hiroshima
        University, Higashi-Hiroshima, Hiroshima 739-8526, Japan},
        S. Aoki\address{Institute of Physics, University of Tsukuba,
        Tsukuba, Ibaraki 305-8571, Japan},
        M. Fukugita\address{Institute for Cosmic Ray Research,
        University of Tokyo, Tanashi, Tokyo 188-8502, Japan},
        S. Hashimoto\address{High Energy Accelerator Research
        Organization (KEK), Tsukuba, Ibaraki 305-0801, Japan},
        K-I. Ishikawa$^{\rm a}$,\\
        N. Ishizuka$^{\rm b,}$\address{Center for Computational Physics,
        University of Tsukuba, Tsukuba, Ibaraki 305-8577, Japan},
        Y. Iwasaki$^{\rm b,e}$, K. Kanaya$^{\rm b,e}$, T. Kaneda$^{\rm b}$,
        S. Kaya$^{\rm d}$, Y. Kuramashi$^{\rm d}$, M. Okawa$^{\rm d}$,\\
        T. Onogi$^{\rm a}$, S. Tominaga$^{\rm d}$,
        A. Ukawa$^{\rm b,e}$, N. Yamada$^{\rm a}$, T. Yoshi\'e$^{\rm b,e}$}

\begin{document}
\begin{abstract}
We report preliminary results of our study of matrix elements of baryon number
violating operators which appear in the low-energy effective
Lagrangian of (SUSY-)Grand Unified Theories. The calculation is performed 
on a $32^{3}\times80$ lattice at $\beta=6.1$ using Wilson fermions 
in the quenched approximation. Our calculation is
independent of details of (SUSY-)GUT models and covers all interesting
decay modes.
\end{abstract}

\maketitle

\section{Introduction}
Proton decay is one of the most exciting predictions of Grand Unified
Theories (GUTs). Experimental effort over the years has pushed 
the lower limit on the partial lifetimes to 
$\tau/B_{p \rightarrow \pi^{0} e^{+}} > 5.5 \times 10^{32}$ years
and $\tau/B_{p \rightarrow K^+ \overline{\nu}} > 1.0 \times 10^{32}$
years at the 90\% confidence level\cite{PDG98}, 
and further improvement is expected from the SuperKamiokande experiment.

A crucial link to relate these numbers to constraints on (SUSY-)GUT models
is the values of hadron matrix elements relevant for proton decay.  
Model calculations suffer from high degree of uncertainty, various
estimations easily differing by a factor ten.

Pioneering lattice QCD studies to remove this source of uncertainty 
were made about ten years ago, 
first combining calculations of the matrix element 
$\langle 0 |O^{\not{B}}| p \rangle$ and soft-pion theorems 
to estimate  $\langle \pi^{0} |O^{\not{B}}| p \rangle$\cite{Hara86,Bowler88},  
and subsequently directly evaluating the matrix element 
$\langle \pi^{0} |O^{\not{B}}| p \rangle$ itself\cite{Gavela89}.

In this article we report preliminary results of our renewed effort 
to determine the matrix elements from first principles of QCD.  
In addition to the use of larger lattice sizes and higher statistics 
to achieve a much better precision, which is made possible through
the increase of computing power, we aim to advance the calculation 
on two fronts:  
(i) calculation of $p\to K$ matrix elements relevant for SUSY-GUT as 
well as those for the $p\to \pi$ mode, for physical values of 
$u$-$d$ and $s$ quark masses and for physical momenta, and 
(ii) evaluation of matrix elements of all dimension-6 baryon number 
violating operators classified according to 
$SU(3)_{C} \times SU(2)_{L} \times U(1)_{Y}$ 
invariance\cite{Weinberg79,Wilczek79}, so as to cover various GUT 
models and decay processes.

\begin{figure}[t]
  \begin{center}
    \psfig{file=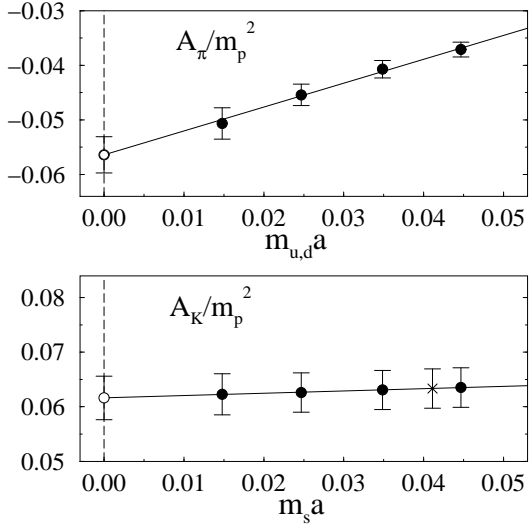,width=7.0cm,angle=270} 
    \vspace*{-3mm}
    \caption{Matrix element relevant for $p\to\pi^0$ and $p\to K^0$ decay 
             normalized by measured proton mass at $\vec{p}=0$ of pseudo 
             scalar meson as a function of 
             $u$-$d$ and $s$ quark mass in lattice units. $u$-$d$ quark 
             mass is taken to the chiral limit for $A_K$.}
    \label{fig:qmass}
  \end{center}
  \vspace*{-12mm}
\end{figure}

\section{Calculational procedure}
Our calculation is carried out in quenched QCD at $\beta$=6.1 with the
Wilson quark action on a $32^{3}\times80$ lattice.  We analyze 100 
configurations for the hopping parameter 
$K$=0.15428, 0.15381, 0.15333, 0.15287. 
The lattice scale fixed by $m_\rho=769$MeV in the chiral limit 
($K_c=0.15499(2)$) equals $a^{-1}=2.56(4)$GeV,   
and the point for strange quark estimated from $m_K/m_\rho=0.648$
is given by $K=0.15304(5)$. All errors are estimated by the single
elimination jackknife procedure.

To calculate the nucleon ($N$) to pseudo scalar (PS) meson matrix element of
a baryon number violating operator $O^{\not{B}}$, we form the ratio,
\begin{eqnarray}
\langle PS | O^{\not{B}} | N \rangle &=&
\frac{\langle 0 | J_{PS} O^{\not{B}} \bar{J}_{N} | 0 \rangle}
     {\langle 0 | J_{N} \bar{J}_{N} | 0 \rangle
      \langle 0 | J_{PS} J_{PS}^{\dagger} | 0 \rangle} \nonumber\\
&\times&
\langle 0 | J_{PS} | PS \rangle
\langle N | \bar{J}_{N} | 0 \rangle,
\end{eqnarray}
where $\langle 0 | J_{PS} | PS \rangle$ and 
$\langle N | \bar{J}_{N} | 0 \rangle$ are extracted from local-local hadron
propagators.   
We fix the nucleon source at $t$=0, PS sink at
$t=32$ and move the operator between them. 
Matrix elements are evaluated for four spatial momenta 
$\vec{p}a$=(0,0,0), (1,0,0), (0,1,0), (0,0,1) in units of 
minimum momentum $ap_{min}=\pi/16$ injected in the PS meson.

We distinguish $u$-$d$ and $s$ quark masses; the former is taken to 
the chiral limit, and the latter interpolated to the physical 
$s$ quark mass in our calculations.
After this procedure, we interpolate the spatial momentum to the
physical value.

Matrix elements are renormalized, with mixing included, by tadpole-improved 
one-loop renormalization factors to the $\overline{{\rm MS}}$ 
scheme\cite{Richards87} calculated at the scale $\mu=1/a$. 

\section{Results}

\begin{figure}[t]
  \begin{center}
    \psfig{file=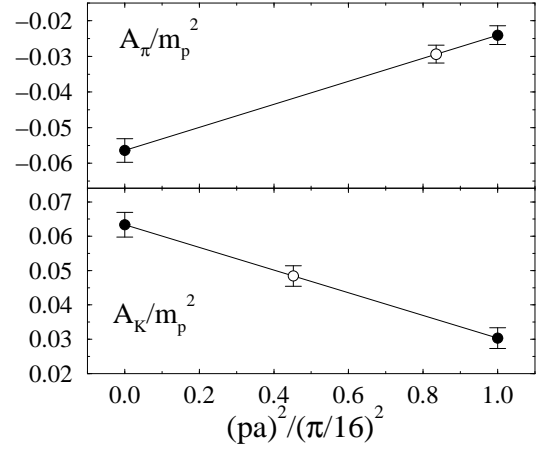,width=7cm,angle=270}
    \vspace*{-10mm}
    \caption{Momentum dependence of $p\to\pi^0$ and $p\to K^0$ matrix 
            elements. Quark masses are taken to the physical point. }
    \label{fig:mom}
  \end{center}
  \vspace*{-12mm}
\end{figure}

Let us define the $p\to\pi^0$ and $p\to K^0$ matrix elements 
$A_\pi$ and $A_K$ by
\begin{equation}
\langle \pi^0(K^{0}) | \epsilon_{ijk} (u^{i}Cd^{j}_{L}(s^{j}_{L}))u^{k}_{R} | 
p \rangle \equiv A_{\pi(K)} N_{R}.
\end{equation}
In Fig.~\ref{fig:qmass} we show our results for these matrix elements
at zero momentum $\vec{p}=0$, normalized by proton mass measured for 
relevant values of $u$-$d$ quark mass.  For $A_\pi$, abscissa represents 
the $u$-$d$ quark mass, while it represents the $s$ quark mass for $A_K$, 
the $u$-$d$ quark mass having been taken to the chiral limit.

Compared to a first calculation of the $A_\pi$ matrix element\cite{Gavela89}, 
whose results are consistent with ours, 
our statistical errors of 4--8\% are improved by about a factor five.  
This allows us to observe that the amplitude exhibits a clear 
decrease of about 40\% from the region of $s$ quark ($m_{u,d}a\approx 0.04$) 
to the chiral limit, 
which was not apparent in results of Ref.~\cite{Gavela89}. 

For the $p\to K^0$ matrix elements $A_K$,  after chiral extrapolation of 
$u$-$d$ quark mass, the dependence on the $s$ quark mass is small. 
The point plotted with a cross 
shows the value interpolated to the physical $s$ quark mass.

In Fig.~\ref{fig:mom} we plot $A_\pi$ and $A_K$ 
as a function of squared momentum after quark masses are 
taken to the physical values. 
We observe a quite significant momentum dependence for both amplitudes, 
necessitating an interpolation for a precise estimate
of their physical values.  The open circles in Fig.~\ref{fig:mom} 
show results of a linear interpolation in $\vec{p}\,^2$ to physical
momentum.

\begin{figure}[t]
  \begin{center}
    \psfig{file=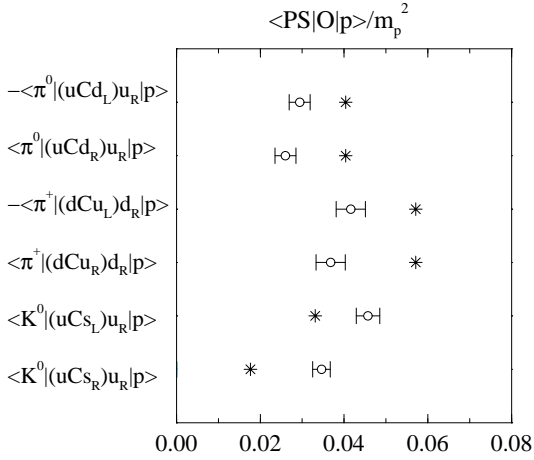,width=7.0cm,angle=270}
    \vspace*{-10mm}
    \caption{Matrix elements for 
             $ p \rightarrow \pi^{0}, \pi^{+}, K^{0}$ decay modes 
             from present work (open circles) compared with 
             predictions of tree-level chiral Lagrangian (asterisks).}
    \label{fig:amp}
  \end{center}
  \vspace*{-12mm}
\end{figure}

Phenomenological analyses of proton decay often employ the soft-pion 
relation\cite{Claudson82,Chadha83}
\begin{equation}
  \langle \pi^{0} | O^{\not{B}} | p \rangle =
  \langle 0 | O^{\not{B}} | p \rangle
  \frac{1}{\sqrt{2}f_{\pi}}(1+F+D),
\end{equation}
to estimate the $p\to\pi^0$ matrix element, where 
$F$ and $D$ are the axial vector matrix elements of 
proton. With our results for physical quark masses and momentum, 
the right hand-side is about four times larger than the left hand-side 
of this relation. A similar discrepancy was observed in Ref.~\cite{Gavela89}. 
Examining the above relation 
for zero pion momentum in the chiral limit, {\it i.e.,} 
the real soft-pion limit, we find that the two 
sides are still discrepant by about a factor two.  The origin of the 
discrepancy is not clear to us at present.

\section{Phenomenology}
We plot our preliminary results for matrix elements normalized by 
proton mass squared relevant 
for $p\to\pi^0, \pi^+$ and $K^0$ decay in Fig.\ref{fig:amp}.
For comparison, predictions from tree-level chiral Lagrangian
with a choice of the parameters $-\alpha=\beta=0.003$ ${\rm GeV}^3$ 
are shown, which represent the smallest values among various 
model estimations. Here $\alpha$ and $\beta$ are defined as  $\langle 0 |
\epsilon_{ijk}(u^{i}Cd^{j}_{L(R)})u^{k}_{R} | p \rangle = \alpha (\beta) N_{R}$.
Our lattice results are even smaller for some of the channels, 
and predictions for the ratio of $p \to\pi^0$ and $p \to K^0$ amplitudes 
are also different.

For non-SUSY minimal SU(5) GUT, the decay width for the $p\to\pi^0$ mode is 
given by\cite{Gavela89}
\begin{equation}
  \Gamma(p \rightarrow \pi^{0} e^{+}) =
  \frac{5\pi}{2} |F_{s}|^{2} \alpha_{5}^{2}(M_{X}) A_\pi^{2}
  \frac{m_{p}}{M_{X}^{4}}.
\end{equation}
Employing the GUT gauge coupling $\alpha_5(M_X)=0.024$ and the short distance
renormalization factor $|F_s|^2=10$ at $\mu=2{\rm GeV}$ as in 
Ref.~\cite{Gavela89}, substituting our preliminary result for 
$A_\pi$ yields 
$\tau/B_{p \rightarrow \pi^{0} e^{+}} = (1.1 \pm 0.1) \times
10^{30} ( \frac{M_{X}}{2.0 \times 10^{14} {\rm GeV}} )^{4}$ years 
where the error is only statistical.
\vspace*{3mm}

This work is supported by the Supercomputer Project No.32 (FY1998)
of High Energy Accelerator Research Organization (KEK),
and also in part by the Grants-in-Aid of the Ministry of 
Education (Nos. 08640404, 09304029, 10640246, 10640248, 10740107, 10740125).
S.K. and S.T. are supported by the JSPS Research Fellowship.

\end{document}